\documentstyle[aps,psfig]{revtex}

\begin{document}

\newcommand{\beq}{\begin{equation}}
\newcommand{\eeq}{\end{equation}}
\newcommand{\A}{\mbox{\bf A}}
\newcommand{\Ss}{\mbox{\bf S}}
\newcommand{\Ha}{\mbox{\cal H}}

\title{Recovery of Protein Structure from Contact Maps}
\author{Michele Vendruscolo$^1$, Edo Kussell$^2$ and Eytan Domany$^1$}
\address{$^1$ Department of Physics of Complex Systems, Weizmann Institute of Sc
ience, Rehovot 76100, Israel}
\address{$^2$ Department of Structural Biology, Weizmann Institute of Sc
ience, Rehovot 76100, Israel}

\address{
\centering{
\medskip\em
{}~\\
\begin{minipage}{17cm}
{}~~~
{\bf Background:}
Prediction of a protein's structure from its amino acid sequence
is a key issue in molecular biology.
While dynamics, performed in the space of
two dimensional contact maps eases the necessary conformational search,
it may also lead to maps that do not correspond to any real three dimensional 
structure. To remedy this, an efficient procedure is needed to 
reconstruct  3D conformations from their contact maps. Such a procedure 
is  relevant also for interpretation of NMR and X-ray scattering
experiments. \\ \\
{\bf Results:}
We present an efficient algorithm to recover the three dimensional structure
of a protein from its contact map representation.
First we show that when a physically realizable map is used as target,
our method generates a structure whose contact map 
is essentially similar to the target. Furthermore, the reconstructed and
original structures are similar up to 
the resolution of 
the contact map representation. 
Next we use non-physical target maps, obtained by corrupting a physical one; 
in this case our method essentially recovers the
underlying physical map and structure. Hence  our 
algorithm will help to   fold proteins, using 
dynamics in the space of contact maps.
Finally we investigate the manner in which the
quality of 
the recovered structure degrades 
when the number of contacts is reduced. \\ \\
{\bf Conclusion:} 
The procedure is capable of assigning quickly and reliably a 3D structure
to a given contact map. It is well suited to be used in parallel
with a dynamics in contact map space to project a contact map onto
its closest physically allowed structural counterpart.
{}~\\
{}~\\
\end{minipage}
}}
\maketitle

\vspace{1cm}

\section{Introduction}

Considerable effort has been devoted to finding ways to predict a protein's
structure from its known amino acid 
sequence $\mbox{\bf A} =(a_{1},a_{2},\dots a_{N})$. 
The {\it contact map} of a protein is a
particularly useful representation of its structure \cite{ls79,hck79}. 
For a protein of 
$N$ residues the contact map is an $N\times N$ matrix \mbox{\bf S}, whose
elements are $S_{i,j}=1$ if residues $i$ and $j$ 
are in contact and $S_{i,j}=0$ otherwise. 
'Contact' can be defined in various ways: for example,
in a recent publication Mirny and Domany (MD) \cite{md96} 
defined contact $S_{i,j}=1$ 
when a pair of heavy (all but hydrogen) atoms, one from amino
acid $i$ and one from $j$, whose distance is less than $4.5\AA $, can be
found. Secondary structures are easily detected from the contact map. Alpha
helices appear as thick bands along the main diagonal, since they involve
contacts between one amino acid and its four successors. The signature of
parallel or anti-parallel beta sheets are thin bands, parallel or
anti-parallel to the main diagonal. On the other hand, the overall tertiary
structure is not easily discerned. The main idea of MD was to use this
representation to perform a search, executed {\it in the space of possible
contact maps} \mbox{\bf S}, for a fixed sequence \mbox{\bf A}, to identify
maps of low ''energy'' $\mbox{$\cal H$}(\mbox{\bf A},\mbox{\bf S})$. They
defined the energy $\mbox{$\cal H$}(\mbox{\bf A},\mbox{\bf S})$ as the
negative logarithm of the probability that structures, 
whose contact map is \mbox{\bf S}, 
occur for a protein with the sequence \mbox{\bf A}; therefore
a map of low energy corresponds to a highly probable structure.

One of the most problematic aspects of their work was the fact that by
performing an unconstrained search in the space of contact maps, i.e. freely
flipping matrix elements from $1$ to $0$ and vice versa, one obtains maps of
very low energy which have no physical meaning, since they do not correspond
to realizable conformations of a polypeptide chain. To overcome this
problem, MD introduced heuristic restrictions on the possible changes 
one is allowed to make
to a
contact map, arguing that if one starts with a
physically realizable map, the moves allowed by these restrictive dynamic
rules will generate maps that also are physically realizable. Even though
their heuristic rules did seem to modify the dynamics in the desired way,
there is no rigorous proof that indeed one is always left in the physical
subspace, there is no clear evidence that the resulting rules are not too
restrictive and, finally, the need to start with a physical fold, copied
from a protein of known structure, may bias the ensuing search and get it
stuck in some local minimum of the energy.

The aim of the present publication is to present a method to overcome these
difficulties. The idea is to provide a test, which can be performed ''on
line'' and in parallel with the dynamics in the space of contact maps, which
will ''project'' any map onto a nearby one which is guaranteed to be in the
subspace of physically realizable maps. That is, for a given target contact
map \mbox{\bf S}, we search for a conformation that a ''string of beads''
can take, such that the contact map \mbox{\bf S}$^{\prime }$ of our string
is similar (or close) to \mbox{\bf S}. Needless to say, the contact map
associated with a string of beads is, by definition, physical.

This particular aim highlights the difference between what we are trying to
accomplish and existing methods \cite{hck79,hw85,bohr93,bohr96,at96,sko97}, 
that use various forms of distance
geometry \cite{ch88} to predict a 3D structure of 
a real protein from its contact
map. Rather than being concerned with obtaining a structure that is close to a
real crystallographic one, we want mainly to check whether a map \mbox{\bf S}
is physically possible or not, and if not - to propose 
some \mbox{\bf S}$^{\prime }$ which {\it is} physical and, 
at the same time, is not too
different from \mbox{\bf S}. The method has to be fast enough to run in
parallel with the search routine
(that uses $\mbox{$\cal H$}(\mbox{\bf A},\mbox{\bf S})$ to identify 
candidate maps \mbox{\bf S} of low energy). Another important requirement is to be
able to recover contacts that do not belong to secondary structure elements
and may be located far from the map's diagonal. Such contacts are important
to nail down the elusive {\it global fold} of the protein. We believe that
the main advantage of performing a dynamic search in the space of contact
maps is the ease with which such contacts can be introduced, whereas
creating them in a molecular dynamic or Monte Carlo procedure of a real
polypeptide chain involves coherent moves of large sections of the molecule
- moves which take a very long time to perform. To make sure that this
advantage is preserved, our method must be able to efficiently find such
conformations, if they are possible, once a new target contact has been
proposed.

We are currently working on combining the  method presented here with dynamics in
the space of contact maps. The results of the combined procedure will be
presented in a future publication.

The plan of the paper is as follows.

\begin{enumerate}
\item  We give a detailed explanation of the method.

\item  We show how it works on native maps 
of proteins with the number of residues ranging from N=56 to N=581. 
Success of the algorithm is measured in terms of
the number of contacts recovered and the root
mean square displacements of the recovered 3D structures from the native ones.

\item  We study the answers given by our algorithm when it faces the task
of finding a structure, using  an unphysical
contact map as its target. As the first test, we added 
and removed contacts
at random
to a physical map and found that
the reconstructed structure did not change by much, i.e.  we still could 
reconstruct the
underlying physical structure.
As a second test we used  the constrained dynamic rules proposed
by MD; starting from an experimental contact map, 
we obtained a new map by a denaturation-renaturation computer
experiment. Since the MD rules are heuristic, this map is not guaranteed
to be physical.
Our reconstruction method projects a non-physical map onto one that is close to it {\it and} physically
allowed.

\item  We discuss the extent to which 
the quality of the structure, obtained from a map, 
gets degraded when the number of given contacts 
is reduced. This issue has considerable importance beyond the scope
of our present study, since
experimental data (disulfide bridge determination, crosslinking studies,
NMR) often are available only for a small number of contacts. 
Clearly the more contacts one has the smaller is the 
number of possible conformations of a chain that are consistent with 
the constraints contained in the contact map. The question
we address is when does this reduction of possible conformations  suffice to define
the corresponding structure with satisfactory accuracy.

\end{enumerate}

\section{Method}

In this work we adopt a widely used definition of contact: 
two amino acids, $a_{i}$ and $a_{j}$ 
are in contact if their distance $d(a_{i},a_{j})$ is less
than a certain threshold $d_{t}$. The distance $d(a_{i},a_{j})$ is defined
as 
\begin{equation}
d(a_{i},a_{j})=|{\bf r}_{i}-{\bf r}_{j}|,
\label{eq:contact}
\end{equation}
where ${\bf r}_{i}$ and ${\bf r}_{j}$ are the coordinates of the $C_{\alpha
} $ atoms of amino acids $i$ and $j$.

The algorithm is divided into two parts. The first part, {\it growth,}
consists of adding one monomer at a time, i.e. a step by step growth of the
chain. The second part, {\it adaptation, } is a refinement of the structure,
obtained as a result of the growth stage, by local moves. In both stages, to
bias the dynamics, we introduce cost functions defined on the basis of the
contact map. Such cost functions contain only geometric constraints, and do
not resemble the true energetics of the polypeptide chain.

\subsection{Growth}

We first describe the growth stage of our procedure.

\subsubsection{Single monomer addition. Here $N_t=5$.}

Suppose we have grown $i-1$ monomers and we want to add point $i$ to the
chain. To place it, as shown in Fig. \ref{fig:growth} 
we generate at random $N_{t}$ trial positions, (typically $N_t=10$),
\begin{equation}
{\bf r}_{i}^{(j)}={\bf r}_{i-1}+{\bf r}^{(j)}\;,
\end{equation}
where $j=1,\ldots ,N_{t}$ and ${\bf r}^{(j)}$ is a vector 
whose direction is selected from a uniform distribution, whereas its
length is distributed normally  with average $r_{a}$ 
and variance $\sigma $. 
Since in our representation monomers identify the $C_{\alpha }$
positions, we took $r_{a}=3.79$ and $\sigma =0.04$. 
We assign a probability $p^{(j)}$ to each trial in the following way. For
each trial point ${\bf r}_{i}^{(j)}$ we calculate the contacts that it has
(see eq. (1)) with the previously positioned 
points ${\bf r}_{1},\ldots ,{\bf r}_{i-1}$. 
Contacts that should be present, according to the given contact
map, are encouraged and contacts that should not be there are discouraged
according to a cost function $E_{g}$ that will be specified below. One out
of the $N_{t}$ trials is chosen according to the probability 
\begin{equation}
p^{(j)}=\frac{e^{-E_{g}^{(j)}/T_{g}}}{Z},  \label{eq:prob}
\end{equation}
where the normalization factor is given by 
\begin{equation}
Z=\sum_{j=1}^{N_{t}}e^{-E_{g}^{(j)}/T_{g}}\;.
\end{equation}
The notation for the cost function $E_{g}$ and for the parameter $T_{g}$
that guide the growth are chosen in the spirit of the Rosenbluth method \cite
{rosenbluth} to suggest their reminiscence to energy and temperature,
respectively.

\subsubsection{Chain growth.}

The step by step growth presented in the previous section optimizes the
position of successive amino acids along the sequence. The main difficulty
in the present method is that the single step of the growing chain has no
information on the contacts that should be realized many steps (or monomers)
ahead. To solve
this problem, we carry out several attempts (typically 10) 
to reconstruct the structure, 
choosing the best one. In practice, this is done as follows.

For each attempt, when position ${\bf r}^{(j)}$ is chosen for monomer $i$
according to Eq. (\ref{eq:prob}), 
its probability is accumulated in the weight 
\begin{equation}
W_{i}=\prod_{k=1}^{i}p_{k}^{(j)}
\label{eq:weight}
\end{equation}
When we have reached the end of the chain we store the weight $W_{N}$. The
trial chain with the highest $W_{N}$ is chosen.

\subsubsection{Cost function.}

The probabilities in Eq. (\ref{eq:prob}) are calculated using the following
cost function : 
\begin{equation}
E_{g}{}^{(j)}=\sum_{k=1}^{i-1}f_{g}(r_{ik}^{(j)})\;,
\end{equation}
where $r_{ik}^{(j)}=|{\bf r}_{i}^{(j)}-{\bf r}_{k}|$, and

\begin{equation}
f_{g}(r_{ik})=d\cdot {\rm a}_g(S_{ik})\cdot \vartheta (d_{t}-r_{ik})\;,
\label{eq:step}
\end{equation}

The enhancing factor $d=i-k$ is introduced to guide the growth towards
contacts that are long ranged along the chain; $\vartheta $ is the Heaviside
step function and the constant ${\rm a}_g$ can take two values; 
${\rm a}_g(S_{ik}=0)\geq 0$ and ${\rm a}_g(S_{ik}=1)\leq 0$. 
That is, when a contact
is identified in the chain, i.e. $r_{ik}<d_{t}$, it is either ''rewarded''
(when the target map has a contact between $i$ and $k$), or penalized. 
In this work we have grown chains with ${\rm a}_g(0)=0$.
In this case, for a given
contact map $S$, the function $f_{g}$ only rewards those contacts that are
realized and should be present. No cost is paid if contacts that are not in
the map are realized by the chain (false positive contacts).
Typically we chose the values ${\rm a}_g(1)=-1.0$ and $T_g=1$.

\subsection{Adaptation}

When we have grown the entire chain of $N$ points, we refine the structure
according to the following scheme. We choose a point $i$ at random and try,
using a crankshaft move, to displace it to ${\bf r}_{i}^{\prime }$. We use a
local cost function $E_{a}^{(i)}:$ 
\begin{equation}
E_{a}^{(i)}=\sum_{k=1}^{i-1}f_{a}(r_{ik}^{\prime })\;,
\end{equation}
where $r_{ik}^{\prime }=|{\bf r}_{i}^{\prime }-{\bf r}_{k}|$, and

\begin{equation}
f_{a}(r_{ik})={\rm a}_a(S_{ik})\cdot \vartheta (d_{t}-r_{ik})\;,
\end{equation}
Note that the enhancing factor $d$ has been omitted, so that $f_{a}$ 
does not favor contacts between monomers that are distant along the chain.
The displacement is accepted with probability $\pi $, according to the
standard Metropolis prescription 
\begin{equation}
\pi =\min (1,\exp (-\Delta E_{a}/T_{a})
\end{equation}
where $\Delta E_{a}$ is the change in the cost function $E_{a}$ induced by
the move and and $T_{a}$ is a temperature-like parameter, used to control
the acceptance ratio of the adaptation scheme.

A key ingredient of our method is annealing \cite{kirkpatrick}. As in all annealing
procedures,
the temperature-like parameter $T_a$ is  decreased slowly during the
simulation to help the system find the ground state in a rugged energy
landscape.

In our method, however, instead of using simulation time as a control parameter on
the temperature, we chose the number $n$ of missing contacts. Two regimes were
roughly distinguished. In the first regime many contacts are missed and
the map is very different from the target one. In the second regime
few contacts are missed, and the map is close to the target. The
parameters ${\rm a}_a$ and $T_{a}$ are  interpolated smoothly between 
values suitable for these two limiting
cases. In the first regime, we strongly favor the recovery of contacts that
should be realized, whereas in the second regime we strongly disfavor contacts
that are realized but should not be present. We set, as shown in Fig. \ref
{fig:annealing} 
\begin{equation}
{\rm a}_a^{(n)}(S)={\rm a}^{f}(S)+[{\rm a}^{i}(S)-{\rm a}^{f}(S)]\sigma (n)\;,
\label{eq:a}
\end{equation}
and 
\begin{equation}
T_{a}^{(n)}=T_{a}^{f}+(T_{a}^{i}-T_{a}^{f})\sigma (n)\;.
\label{eq:t}
\end{equation}
The function $\sigma (n)$ interpolates between the initial value ${\rm a}^{i}$ and
the final value ${\rm a}^{f}$, 
\begin{equation}
\sigma (n)=\frac{2}{1+e^{-\alpha_g n}}-1\;.
\end{equation}
By choosing ${\rm a}^{i}$, ${\rm a}^{f}$, 
$T_{a}^{i}$, $T_{a}^{f}$ and $\alpha_g $ we
define the two regimes, far from  and close to the target map. 

\subsection{Chirality}

A contact map contains no information about chirality. When an overall
structure is reconstructed, the mirror image conformation is equally
legitimate, having the same contact map. Since existing proteins do have a
definite chirality, we are allowed to supply this information.

The $C_{\alpha }-C_{\alpha }$ distance constraints are loose enough to allow
a local refinement of the reconstructed structure, with no loss in our
geometrical cost function. Alpha helices can be detected as a thick band
along the main diagonal of a contact map. A preliminary scan of the map
identifies the sections that should be reconstructed as alpha helices. Next,
we push the $C_{\alpha }$'s in the alpha helices to positions that give the
correct chirality, which is formally defined as the normalized triple
product 
\begin{equation}
c_{i}=\frac{{\bf v}_{i}\times {\bf v}_{i+1}\cdot {\bf v}_{i+2}}
{|{\bf v}_{i}| \cdot |{\bf v}_{i+1}| \cdot |{\bf v}_{i+2}|}\;,
\end{equation}
where ${\bf v}_{i}={\bf r}_{i}-{\bf r}_{i-1}$.

In a typical alpha helix, $c_{i}=c_{o}=.778$ \cite{bohr96}. To refine the
chirality of the preliminary chain that was obtained from the map by growth and
adaptation as described above, we perform an additional Monte Carlo procedure.
This procedure uses as ''energy'' a function that strongly
favors the value quoted above for $c$: 
\begin{equation}
E_{c}=a_{c}\left[ \frac{2}{1+exp\left[ -\alpha _{c}(c-c_{o})^{2}\right] }
-1\right] \;,  \label{eq:chirality}
\end{equation}
Since our Monte Carlo moves do not conserve the $C_{\alpha }-C_{\alpha }$ bond
length, we added a term $E_{b}$ to the
energy function 
\begin{equation}
E_{b}=a_{b}\left[ \frac{2}
{1+exp\left[ -\alpha _{b}(r-r_{a})^{2}\right] } -1\right] \;.  
\label{eq:bond}
\end{equation}
At each step a monomer $i$ is selected randomly and its position displaced
to 
\begin{equation}
{\bf r}_{i}^{\prime }={\bf r}_{i}+\delta \;,
\end{equation}
where $\delta $ is a small random vector. The total variation in the cost
function, $E_{a}+E_{c}$, is evaluated with 
\begin{equation}
c=\frac{c_{i-1}+c_{i}+c_{i+1}}{3}
\end{equation}
used in Eq. (\ref{eq:chirality}).

Growth and adaptation yield a particular recovered structure, ${\bf C}$. We first
create $\bar {\bf C}$,  the mirror image of ${\bf C}$ and use both structures as
initial states for the final refinement procedure.\\
Usually either ${\bf C}$ or $\bar {\bf C}$ evolves to  a structure with the correct value of the
average chirality rather  quickly  by our Monte Carlo process,	while the mirror image
does not, due to 
the lower compatibility of the latter structure
with the correct chirality.

\subsection{Remarks.}

We devote the rest of this section to the discussion of alternative
strategies that we have tried.
Some of these  may prove to be useful in 
future applications for more complex problems, but
we have found that they are non necessary
for the specific task dealt with in this work.
We present these experiments since they underscore some non trivial aspects
of the problem.

{\it Adaptation alone;}
It is interesting to note that for short chains ( $N<200)$ we can skip the
growth stage;
starting from a
random structure, the adaptation procedure alone suffices to recover the
correct set of contacts. The computer time needed for recovery increases,
however, very fast with $N$.  Since we are interested in recovering the
structure in as short a time as possible, growth must be 
used, especially for long chains. We found that starting the
adaptation stage from a grown (versus random) initial chain speeds up the
procedure by a factor of about ten for proteins of length $N \simeq 100$. 
Moreover, for longer chains ($200<N<1000$), 
the cost function landscape is rougher and reconstruction by adaptation
alone becomes unfeasible.

{\it Piecewise growth:}
The importance of local contacts (i.e. contacts that involve amino acids
nearby along the chain) versus non local ones has been discussed recently
in the literature \cite{ags95,gg95}. In these works evidence is given in
support of the idea that non local interactions
are decisive to stabilize the folded structure.
A long standing alternative hypothesis \cite{as75} is that the folded structure
is  stabilized mainly by local interactions. We can test in the present work 
whether
the {\it purely geometrical} (versus energetic) part of the reconstruction 
can or cannot be helped much by emphasizing the role of local contacts. To this end
we  used secondary structure elements
as guidelines for the step by step growth.
To implement this kind of growth instead of growing the entire chain of $N$
amino acids, a section of $M$ steps is built,
with $M$ ranging from 4 to 10 to match
the size of a turn in an alpha helix or in a beta sheet. 
A set of sections is generated and the one with the best weight,
according to Eq. (\ref{eq:weight}), is chosen.
Consistently with the findings in Refs. \cite{ags95,gg95},
we found that this secondary structure driven growth is not helping
much the recovery.

A related idea is to
optimize the relative positions of
successive secondary structure elements. 
To realize this idea we have tried the following method.
Similarly as above, sections of chain of $M$ steps are grown,
but now $M$ is chosen randomly from 20 to 50. In this way we explore
the space forward on the length scale of secondary structures to hook 
important contacts, i.e. those that fix the  positions of secondary
structures relative to each other. This scheme biases the growth 
to build a bridge to the next
important contact, which usually is either inside a secondary structure or
between different secondary structures. This method also allow to go back if
too many mistakes are detected. As was the case for the
previous attempted method, our experience suggests that
this forward exploration is not necessary for solving the present task.

For multidomain proteins we  tried to grow one domain at a time and then
refining the structure by an adaptation cycle. The
overall results were, however, only slightly affected.

An alternative idea that we tried is 
to bias the growth  towards reaching
a particular 'fixed point" \cite{v97}. For example, if it is known that two amino
acids $i$ and $j=i+k$ should be in contact then it is possible to bias the
formation of a loop of length $k$. This method is well suited for very
sparse contact maps, where it is easy to identify target points for the
growing chain. We have verified that in dense maps the reconstruction speed
is not increased by this scheme, due to the cumbersome identification of the
target points.

{\it Different cost functions:}
As mentioned above, we have used ${\rm a}_g(0)=0$.
In general,
this fact could lead to an overcompactification of the final structure. To
assign unfavorable weight to false positive contacts,
we should set ${\rm a}_g(0)>0$.
This would introduce, however, frustration to the growth process,
since it is guided by positive and negative energies. We discuss here
the results of a possible 
way that we have tested to bias the growth away from conformations that contain
spurious contacts that are not present in the map. We have 
assigned a positive cost
${\rm a}_g(0)>0$ to generating a spurious contact $(i,j)$ if the closest
existing contact (as measured on the map) is more than a distance of $R$ units a
way,
e.g.
\begin{equation}
\min_{(h,k)}\sqrt{(i-h)^{2}+(j-k)^{2}}>R\;,
\end{equation}
where $(h,k)$ run over all the existing contacts $S_{h,k}=1$ in the given map. 
For the proteins
we have analyzed (see Table \ref{tab:test}), we have extensively
scanned possible values for $R$ and ${\rm a}_g(0)$. We  found
that that there is a strongly frustrated regime, for small $R$ and
large ${\rm a}_g(0)$, where reconstruction is hindered;
and a weakly frustrated regime, for large $R$ and
small ${\rm a}_g(0)$, where the efficiency of the reconstruction
is only slightly improved with respect to using
${\rm a}_g(0)=0$. The intermediate regime, (typically
$R=5-10$ and ${\rm a}_g(0)=0.1-0.01$ for the values
${\rm a}_g(1)=-1.0$ and $T_g=1$ given above),
may prove to be useful for proteins longer than those tested
is the present work.

As for the functional form of the cost function, another possible choice, 
following Bohr {\em et al.} \cite{bohr93,bohr96}, 
is to smooth the step function that defines a contact with a sigmoid: 
Again, we did not find this necessary to achieve fast reconstruction.

In principle it would be possible to add to the function $f_g$ of (\ref{eq:step})
a hard core
repulsion 
\begin{equation}
h(r)= \sigma_0(r-r_0)^{-\alpha} \;,
\end{equation}
to try to overcome a general problem which arises when working with distance
inequalities: an
overcompactification of the globule, as measured, e.g. by the gyration
radius. In practice, a good recovery prevents automatically the overlap
between alpha carbons and the addition of such a term is not necessary.

\section{Experimental contact maps.}

In this section we present results about the reconstruction of experimental
contact maps as taken from PDB. Since our purpose,
as explained in the Introduction, is to use the
reconstruction in connection with dynamics, we chose $d_t=9 \AA$ to
obtain the most faithful representation of the energy of the protein \cite
{md96}. Such a threshold is determined by the requirement that the average
number of $C_{\alpha}-C_{\alpha}$ contacts for each amino acid is roughly
equal to the respective numbers obtained with the all-atom definition of
contacts.

Two dissimilarity measures
between structures are widely used. The most commonly adopted is
the root mean square distance $D$ 
\begin{equation}
D = \sqrt{\frac{1}{N} \sum_{i=1}^N ( {\bf r}_i - {\bf r}_i^{\prime})^2} \; ,
\end{equation}
where one structure is translated and rotated to get a minimal $D$.
The standard procedure, described in Ref. \cite{kabsch}, was used. Another
possible choice is the distance $D^{\prime}$ 
\begin{equation}
D^{\prime} = \sqrt{\frac{1}{N^2} \sum_{i,j=1}^N ( r_{i,j} -
r_{i,j}^{\prime})^2} \; ,
\end{equation}

The dissimilarity measure between contact maps is defined as 
the Hamming distance
\begin{equation}
D^{{\rm map}} = \sum_{j>i} | S_{ij} - S_{ij}^{\prime} | \;,
\end{equation}
which counts the number of mismatches between maps 
${\bf S}$ and ${\bf S^{\prime}}$.

For several proteins, we present 
in Fig. \ref {fig:distance_length}
the distances $D$ and $D^{\prime}$ plotted
vs the chain length $N$.
The proteins considered, (with their respective lengths $N$ and number
of contacts $N_c$) are  

The
values of $D$ and $D^\prime$ presented in Fig. \ref{fig:distance_length} 
were obtained by averaging over 100 reconstruction
runs for chains up to $N=223$ and over 10 runs for longer chains. 
Error bars represent the variances as obtained from the corresponding sets of runs,
as shown, for example,  for the proteins 
6pti bovine pancreatic trypsin inhibitor)
and 1trmA (rat trypsin, chain A)
in Fig. \ref{fig:runs}.

In Fig. \ref{fig:6pti_map} we show the contact map for the protein 6pti,
$N=56$, as taken from PDB, that was used as a target to construct
a chain. The contact map of a typical reconstructed chain is also shown.
In this particular case none of the 342 original contacts were missed and only
two false positive contacts were added. These are close to clusters of correct
contacts, indicating slight local differences with the crystallographic 
structure.
The distances recorded in this case were $D^{\prime}=1.06$ and $D=1.56$.

In Fig. \ref{fig:1trma_map} we show similar results 
for the larger protein 1trmA,
with $N=223$ and 1595 contacts. For
clarity we have separated the experimental contact map from the
reconstructed one. In the particular case shown there are 9 missing
contacts and 84 false positives,
and the corresponding distances are $D^{\prime}=1.34$ and $D=1.59$.
On average, in the 100 runs, 6 contacts were missed and 75 false positives
were spuriously added. As in the case of 6pti, wrong contacts are 
mostly neighboring correct ones.
Averages distances are $\langle D^{\prime} \rangle =1.3 \pm 0.1 \AA$
and $\langle D \rangle =1.6 \pm 0.2 \AA$ (see also Fig. \ref{fig:runs}).

Using the
distances $D$ and $D^{\prime}$ to assess the quality of
our results is misleading, since we are searching only for a chain that {\it
reproduces the contacts} of a given map, whereas  $D$ and $D^{\prime}$
measure similarity between {\it structures}. Information, which is all-important
to obtain low values for $D$ and $D^{\prime}$, such as the
positions of amino acids that belong to loops, or slight rotations of secondary
structures, is not contained at all in the map.  
For example, for the two-domain protein 1pii (phosphoribosylanthranilate isomerase),
which has the largest distance in Fig. \ref{fig:distance_length}, only 
{\it two out of 3070 contacts were missed}, on the average,  in the 10 reconstruction
runs. However, changes in the relative orientations of the two domains lead
to large distances.
In fact, the target native maps  
were nearly perfectly reconstructed for all proteins tested.

We turn now to estimate the range of expected values for $D$ and $D^{\prime}$.
The lower limit of our resolution for the chain, imposed by
the geometrical constraints 
contained in the contact map, is about $1 \AA$. 
To support this statement, we present the results of the following test.
We subjected the native maps of 6pti, 1acx (actinoxanthin) 
and 1trmA from PDB and we 
subjected them to an adaptation cycle at low $T_a$.
No native contacts were 
lost and no spurious ones were generated
throughout the simulation, 	even though the structure (i.e. the positions of
the beads) did vary; 
the most probable
value for the distance $D^{\prime}$ between the generated structures was found
to be around $1 \AA$.
This result clearly indicates the extent to which the contact map
representation does not allow to nail down one specific structure
to arbitrary precision.
This ambiguity is compatible with the usual experimental 
resolution of PDB structures and hence
the contact map representation is useful.
Moreover, from low temperature flash photolysis experiments \cite{frau75},
X rays diffraction results analysis \cite{frau88} and molecular dynamics
simulations \cite{ek87},
the native fold of a protein is believed to consist of a set of
conformational substates rather than of a unique structure \cite{fsw91}.

The upper limit of the range of expected distances in our reconstruction is that
between two completely unrelated
structures, which can be as large as $15 \AA$. 

The conclusion of our studies is that our method produces, using
a native contact map as target, a structure whose contact map is in nearly
perfect agreement with the target. 
Furthermore, the distance of this reconstructed chain  from the native structure
is quite close to the resolution that can be obtained from the information contained 
in contact maps.

\section{Non physical contact maps}
As stated in the Introduction, our
main purpose is to develop a strategy to construct a three dimensional
structure, starting from a given set of contacts, even if these
contacts are not physical, i.e. not compatible with any 
conformation allowed by a chain's topology.	 In such a case we require 
our procedure to yield a chain whose conformation is 
as ``close'' as possible to the contact map
we started with. The exact measure of such closeness 
depends on the source of the non-physicality, as will be demonstrated in 
two examples described below.

Our first examples of non physical contact maps 
were obtained by randomizing a native contact
map; this was done by flipping $M$ randomly chosen entries.
Contacts between consecutive amino acids (neighbors along the chain) were conserved.

A typical contact map with noise is shown 
in Fig. \ref{fig:6pti_noise}. The protein is 6pti, whose  map 
has 342 contacts. We show the native map and the target
map obtained by flipping at random $M=100$ entries of the native map, together with 
the map  produced by our method.
The most important conclusion that can be drawn from Fig. \ref{fig:6pti_noise}
is that isolated non-physical contacts are identified as such  and ignored
and the underlying physical contact map is recovered.

The dependence of this recovery on the noise level is
shown in Fig. \ref{fig:6pti_dis_noise}, where we present
the average distance of 
the final structure from the uncorrupted 6pti map, for various values of $M$.
Averages were taken over 10 reconstruction runs.
The distance for totally unrelated structures 
for 6pti is around $8 \AA$. 
It is remarkable that up to $M<200$ a fair resemblance to the experimental
structure is still found.

The results of a similar study of a longer protein,
1trmA, with $M=400$ flipped contacts, is shown in Fig. \ref
{fig:1trma_no400}. For the particular case shown, distances to the PDB
structure are $D^{\prime}=2.1$ \AA and $D=2.4$ \AA. The PDB map has 1595
contacts.

The family of possibly
non physical contact maps, which is most relevant
to our program, is  produced by using the heuristic constrained
dynamic rules in contact map space that were introduced by MD.
Following them, we started with a native map of 6pti; when 
using a threshold $d_t = 8.5 \AA$, the map derived 
from the PDB structure has 289 contacts,
represented by open circles above the diagonal of 
Fig. \ref{fig:6pti_cdyn}. 
We have recomputed the energy parameters for
the present definition of contacts (which involves $C_{\alpha}$ atoms only)
and for a threshold of $8.5 \AA$ \cite{vd97}. 
The energy of the native 6pti map, obtained in this way, is 36.81.
This contact map is subjected to repeated denaturation-renaturation
cycles, using the constrained dynamics introduced by MD.
We first heat the protein, inducing its unfolding, which is signaled
by melting of secondary structure elements in the contact map.
For moderate temperature shocks the protein is generally able to refold
upon annealing \cite{md96}. 

In this work, we add a second step to this experiment, by subjecting the contact
map obtained by constrained dynamics to our reconstruction procedure.
To discuss in some detail the result of this combined scheme,
we introduce three classes of contacts: we denote by A the contacts
present in $\mbox{\bf S}_A$, the experimental contact map of protein 
6pti (native contacts);
by B the contacts present in $\mbox{\bf S}_B$, 
the contact map obtained by constrained dynamics in contact map space,
starting from $\mbox{\bf S}_A$;
and by C the contacts in $\mbox{\bf S}_C$,
the reconstructed contact map obtained using $\mbox{\bf S}_B$ as a target.
We present in Table \ref{tab:cdyn} the total number of contacts
in each class and the number of contact in common between two classes.
Map $\mbox{\bf S}_B$ has 255 contacts, 215 of which are in common
with map $\mbox{\bf S}_A$. This difference is due to 
74 missing contacts and 40 spurious ones (see also Fig. \ref{fig:6pti_cdyn},
above the diagonal). The energy of $\mbox{\bf S}_B$ is 13.35.
The reconstructed map $\mbox{\bf S}_C$ has 310 contacts, 251 of which
are in common with the target map $\mbox{\bf S}_B$. The difference
arises from 4 missing contacts and 61 false positives.
This reconstruction score is significantly larger than those typically
obtained by applying directly our reconstruction procedure to native maps
of proteins of similar size (see e.g. discussion about Fig. \ref{fig:6pti_map}).
This suggests, although without proving it,
that the first step of the experiment, when we apply
the rules of constrained dynamics introduced by MD
is not guaranteed to yield a physically realizable map.
However, from a closer inspection we can derive an overall consistency
argument that implies that the MD rules do not drive the system very far away
from the physical region in contact map space.
The map $\mbox{\bf S}_C$ and the native map $\mbox{\bf S}_A$, 
have 249 common contacts. $\mbox{\bf S}_C$ has 40 missing and 
61 false positives with respect to $\mbox{\bf S}_A$.
Nevertheless, the distances in the 3D structures are 
$D^{\prime}=2.14$ and $D=2.97$ respectively, indicating a 
rather successful refolding.

From these results we argue that MD rules alone are possibly not enough
restrictive to keep the trajectory of the system in the
physical region of the contact map space during a denaturation-renaturation
experiment.
This problem can be corrected by the reconstruction procedure
discussed in this work.
Our procedure projects the contact map obtained by MD rules onto
a contact map which is admissible by construction.
Rather consistently, the projected map is quite close to the
target one.

\section{Reducing the number of contacts}
In this section we address a very important issue; the effect of reducing the
number of contacts on the accuracy of the reconstructed structure. Even though
resolving this problem is not essential for our goals, its resolution is an
interesting spinoff obtained from our algorithm. The question is relevant
to a number of problem areas where contacts are of importance, such as	protein
structure determination from NMR 
data \cite{w86}, studies of DNA and crosslinked polymers. The latter
are known to undergo a vulcanization transition from
a liquid phase to a frozen amorphous phase if the number
of contacts exceeds a critical value \cite{vulcan}.
The stochastic reconstruction method described in this work is rather general
and potentially applicable to these systems as well.

In real proteins the number of contact scales with the
length $N$ of the chains, as shown in Fig. \ref{fig:scaling} for a
representative set of 246 proteins taken from PDB. 
Fitting the data with a single power law
\begin{equation}
M = aN^{\nu} \; ,
\end{equation}
yields best fit for $\nu=1.07$ but the data are also
compatible with  linear scaling  (e.g. $\nu=1$ ) 
as well as with a combination of linear scaling and surface corrections 
\begin{equation}
M =aN + bN^{2/3} \; .
\end{equation}
All three fits are are shown in the figure and are nearly indistinguishable 
on the scale used.
Fig. \ref{fig:scaling} was obtained using the definition of contacts
as given by Eq. (\ref{eq:contact}) with a threshold of
$d_t =9 \AA$. In the range $5 \AA < d_t < 9 \AA$, only the prefactor $a$
changes, while the exponent $\nu$ remains the same.
This result holds also for the MD definition of a contact.
It has been proposed that in order to have a
compact structure, the minimum number of contacts of a random heteropolymer 
should scale
linearly with $N$ \cite{cs97,gs94} or with $N/\ln N$ \cite{bt96}. These
findings suggest that in proteins the number of contacts required to
determine the native fold also cannot scale 
with a power that is much less than linear with $N$. The
relevant issue, to which considerable 
effort has been recently devoted\cite{at96,sko97}, 
concerns  how small can be the prefactor $a$, in order to 
achieve reasonable
reconstruction of protein structure from incomplete experimental contact
maps. We address this point by analyzing the feasibility of the
reconstruction as the threshold $d_t$ is decreased. The smaller $d_t$, the
smaller is the number of contacts present in the contact map. 

We now present detailed studies of the protein  1acx with $N=108$. For $d_t=9 \AA$ the 
number of contacts was 652; for $d_t=6$ this number becomes 253 and 
154 for $d_t=5$. 
Note that the optimal parameters used for annealing depend on
$d_t$; 
for $d_t=5 \AA$, for example the values 
${\rm a}^f(1)=-5$, 
${\rm a}^i(1)=-20$, ${\rm a}^f(0)=0.1$ and  ${\rm a}^i(0)=0.5$ were used.

In all cases our method produced chains whose contact maps were
in nearly perfect agreement with the respective target maps (deviating by 
one or two spurious
contacts). The distances of the corresponding structures from the
native one are, however, very different as $d_t$ decreases. As shown on 
Fig. \ref{fig:constr}, the values of the average distances $D^\prime, D$ 
(obtained from 100 runs
for each $d_t$) increase, from  $ 1 - 2\AA$ for $d_t=9\AA$ to $5 - 8 \AA$ for
$d_t = 5\AA$. 
This striking increase of $D$ with decreasing $d_t$ shows that
even when the target contact map is essentially perfectly recovered, 
the corresponding structure can be very different from the
true one. This suggest that for low values of $d_t$ 
more information than what is contained in the
contact map should be provided to get acceptable resemblance to the
experimental structure.

\section{Summary and Conclusions}

We presented a stochastic method to derive a 3D structure
from a contact map representation.
We have shown that for physically realizable target maps  
our method is very fast and reliable to find a chain conformation whose contact map
is nearly identical to the target.
Moreover, a the method is able to find a good candidate structure even when
the target map has been corrupted with non-physical contacts.

The information contained in the contact map derived from 
an experimental structure allow us to reconstruct
a conformation, which is relatively close to that of the original structure, as
already observed by Havel {\em et al.} \cite{hck79}.
There is, however, an intrinsic limit in the resolution of a contact map.
When  a threshold of $9 \AA$ between $C_\alpha$ atoms is used to define contact,
the distance between two typical structures, that are both
compatible with the contact map, is of the order of $1 \AA$.
Some recent work  focused on the identification of supplemental
information needed to improve the resolution of the reconstruction
\cite{bohr93,bohr96,at96,sko97}. 
Usually, energy terms are introduced
to favor secondary structures, side chain burial, hydrogen bonding	and
sulfur bridges. An example for such terms is the chirality 
term used in the present work.
Although these issues are of considerable importance for 
experimental structure determination,
we consider the problem from a different perspective.

The contact map representation is intimately connected with the
pairwise parametrization of the energy.
We are currently studying the 
dynamics in contact map space, controlled by such a
pairwise energy function.
We believe that such a study will reveal whether  the 
contact map representation, together with the assumptions implicit in working with
a  pairwise, contact-based approximation for the energy,  suffice
to single out the native state of a protein.

This research was supported by grants from the Minerva Foundation,
the Germany-Israel Science Foundation (GIF) and by a grant from the 
Israeli Ministry of Science. E. D. thanks the members of the Laboratoire
de Physique Statistique at the Ecole Normale Superieure, Paris for their
hospitality. E.K. thanks the K. Kupcinet fellowship program for partial
support of his stay as a summer student at the Weizmann Institute.



\begin{figure}
\caption{Single monomer addition.}
\label{fig:growth}
\end{figure}

\begin{figure}
\caption{Annealing functions for the parameters used in adaptation.
See Eqs. (\protect\ref{eq:a}) and (\protect\ref{eq:t}).}
\label{fig:annealing}
\end{figure}

\begin{figure}
\caption{Average distances
$\langle D \rangle$ (lower) and $\langle D^{\prime} \rangle$ (upper) 
vs chain length $N$ for the proteins listed in Table \protect\ref {tab:test}.}
\label{fig:distance_length}
\end{figure}

\begin{figure}
\caption{Histograms of the distances $D^{\prime}$ (dots) and $D$ (open circles)
for the 100 runs used to test the reconstruction
procedure. Data are presented for proteins 6pti (left) and 1trmA (right).}
\label{fig:runs}
\end{figure}

\begin{figure}
\caption{Contact maps for protein 6pti for a threshold $d_t = 9\AA$. 
Dots are the PDB data,
open circles the output of the reconstruction procedure.
None of the taget contacts is missed and two spurious ones are added.}
\label{fig:6pti_map}
\end{figure}

\begin{figure}
\caption{Contact map of protein 1trmA. 
Experimental contact map (above diagonal)
and reconstructed one (below diagonal).}
\label{fig:1trma_map}
\end{figure}

\begin{figure}
\caption{Above diagonal:  target map
(crosses) obtained by randomizing the underlying physical map (dots) 
of the protein 6pti. Below diagonal: reconstructed contact map (open circles)
obtained by using the noise-corrupted  map as target. The physical map
is also shown (dots).}
\label{fig:6pti_noise}
\end{figure}

\begin{figure}
\caption{Average distances $D$ (open circles) and $D^{\prime}$ (dots)
vs noise $M$ for 6pti.}
\label{fig:6pti_dis_noise}
\end{figure}

\begin{figure}
\caption{Above diagonal:   reference map
(crosses) obtained by randomizing the underlying physical map (dots) 
of protein 1trmA.  Below diagonal: reconstructed contact map (open circles)
obtained using the noise-corrupted map as target.}
\label{fig:1trma_no400}
\end{figure}

\begin{figure}
\caption{Reconstruction of a contact map obtained by the 
constrained dynamics introduced by MD.
Above diagonal: reference map (dots), 
obtained by a denaturation-renaturation cycle
starting from the experimental map of protein 6pti (open circles).
Below diagonal: reconstructed contact map (crosses), obtained by using
as a target the contact map (dots) that was obtained by constrained dynamics.}
\label{fig:6pti_cdyn}
\end{figure}

\begin{figure}
\caption{Scaling of the number $M$ of contacts with the length $N$ of the
proteins. Data refer to 246 proteins taken from PDB, and to
a threshold $d_t=9 \AA$.}
\label{fig:scaling}
\end{figure}

\begin{figure}
\caption{Average distances $\langle D^{\prime} \rangle$ (dots) 
and $\langle D \rangle$ (open circles)
as a function of the threshold $d_t$ for protein 1acx.}
\label{fig:constr}
\end{figure}


\newpage
\begin{table}
\caption{List of PDB proteins used to test the reconstruction procedure.}
\label{tab:test}
\end{table}

\begin{table}
\caption{Number of contacts in the classes A, B and C explained in the text.
Also the number of common contacts between classes is reported.}
\label{tab:cdyn}
\end{table}

\newpage
\begin{figure}
\centerline{\psfig{figure=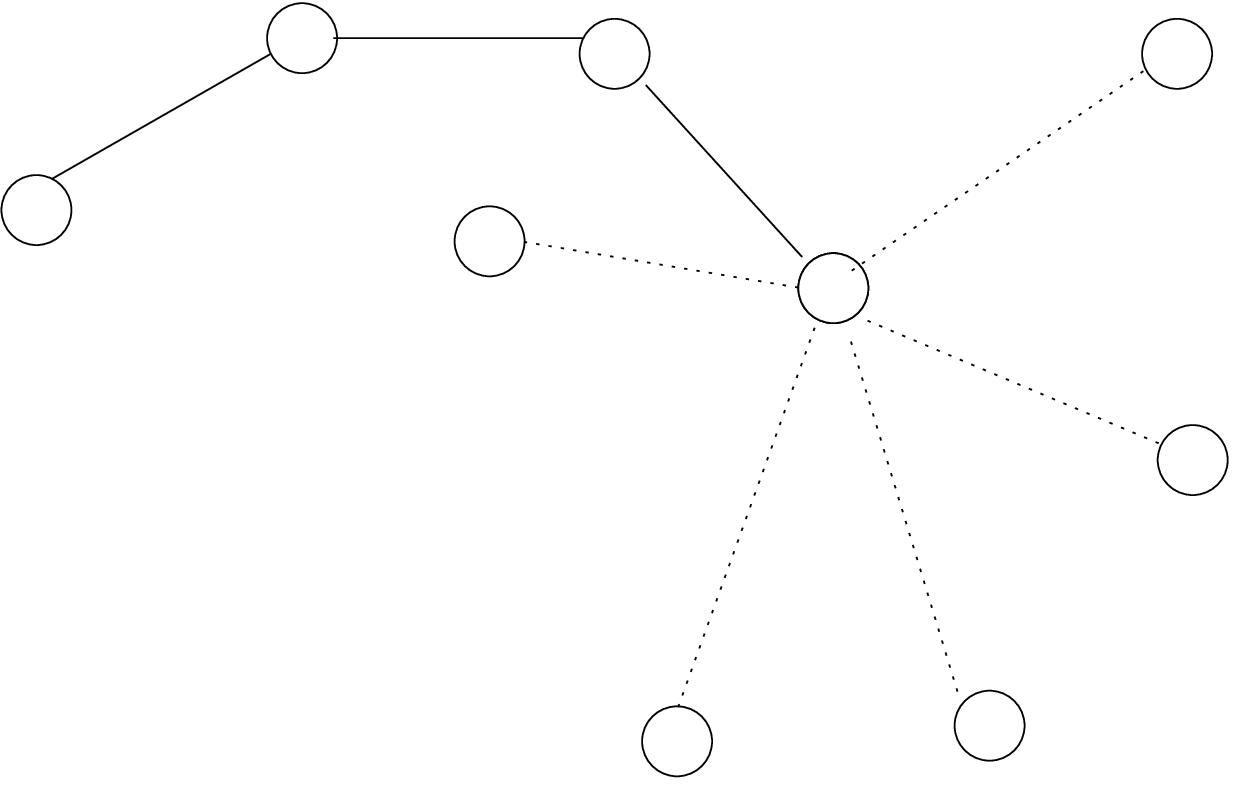,height=8.0cm,angle=270}}
\end{figure}
\newpage
\begin{figure}
\centerline{\psfig{figure=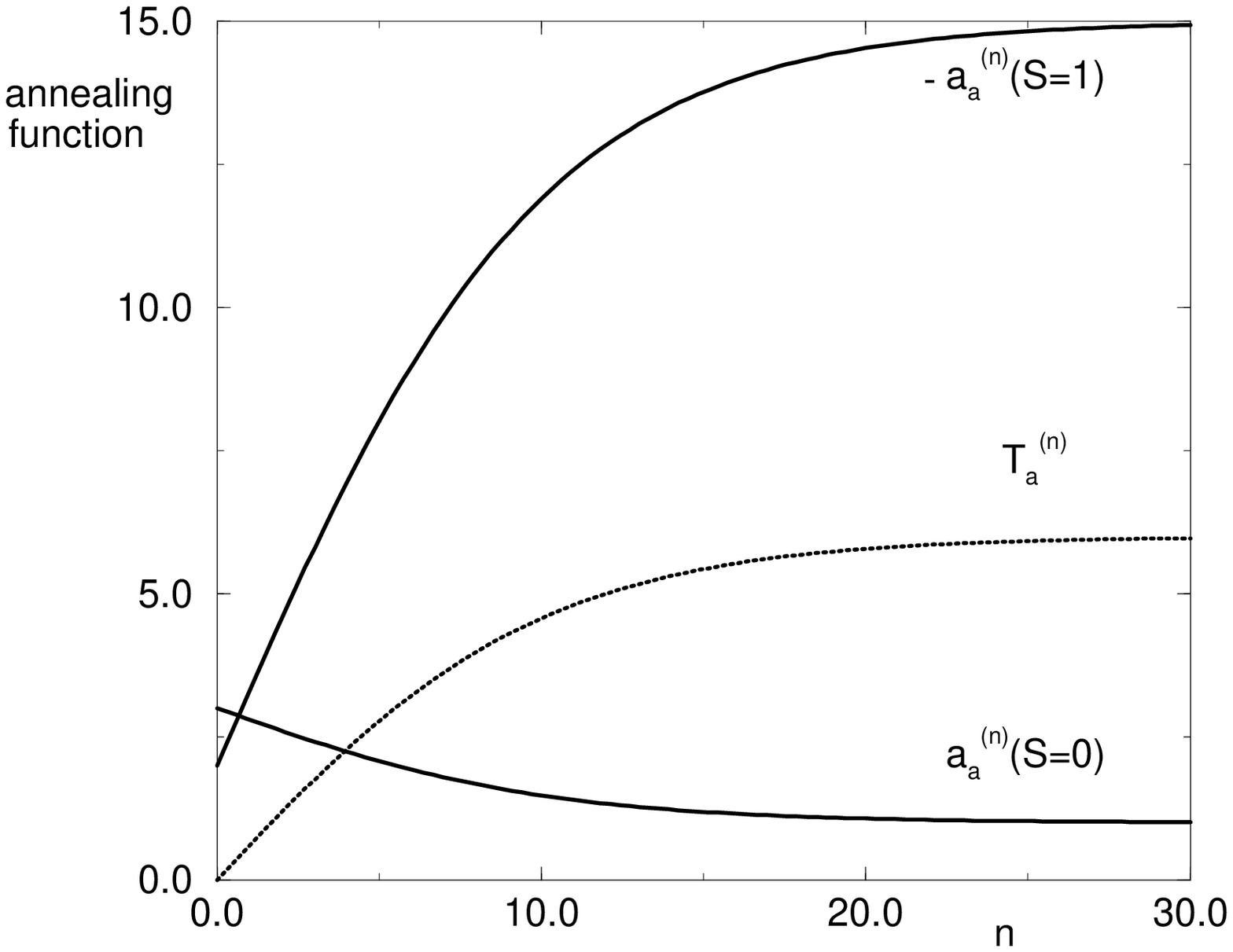,height=15.0cm,angle=0}}
\end{figure}
\newpage
\begin{figure}
\centerline{\psfig{figure=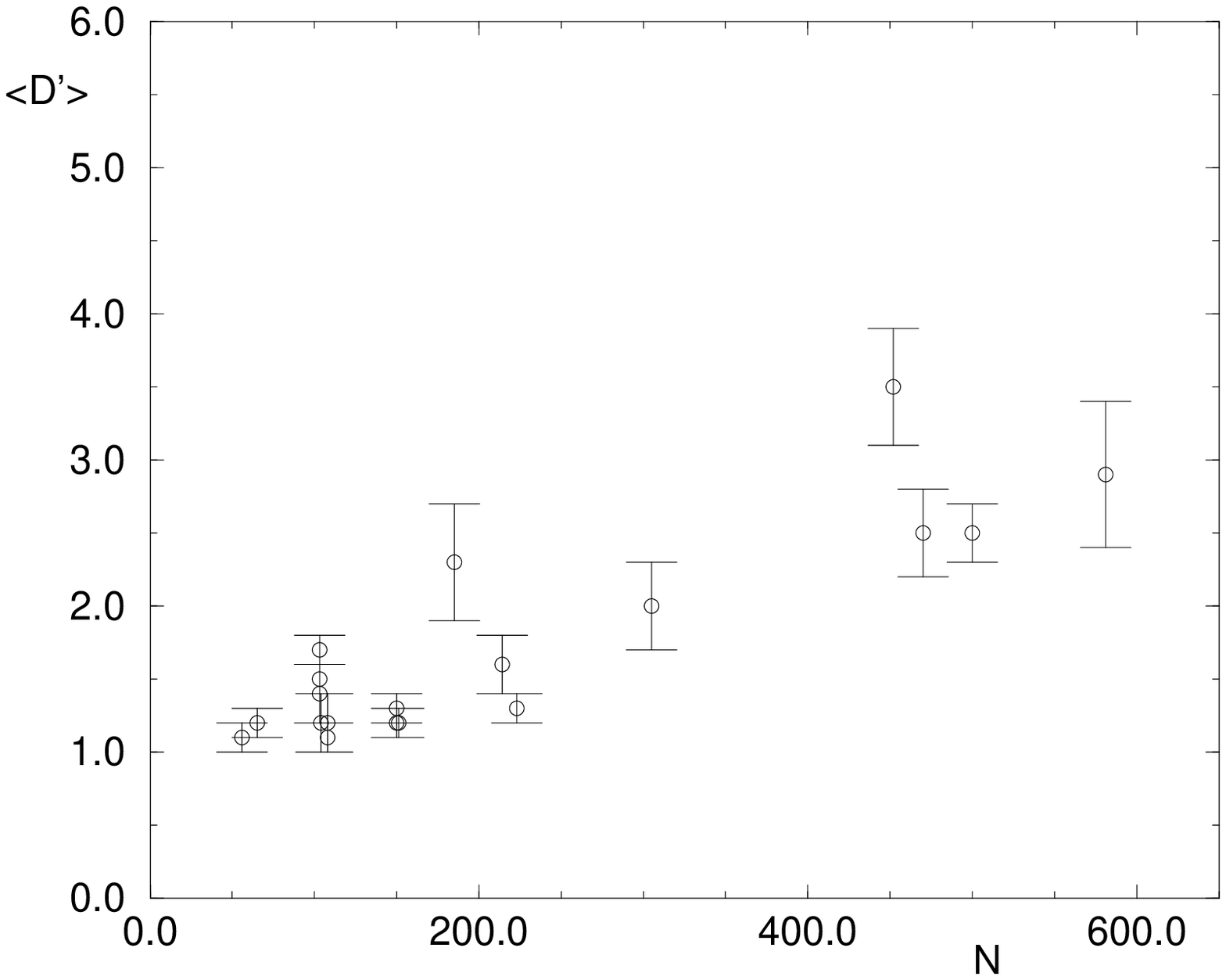,height=9.5cm,angle=0}}
\centerline{\psfig{figure=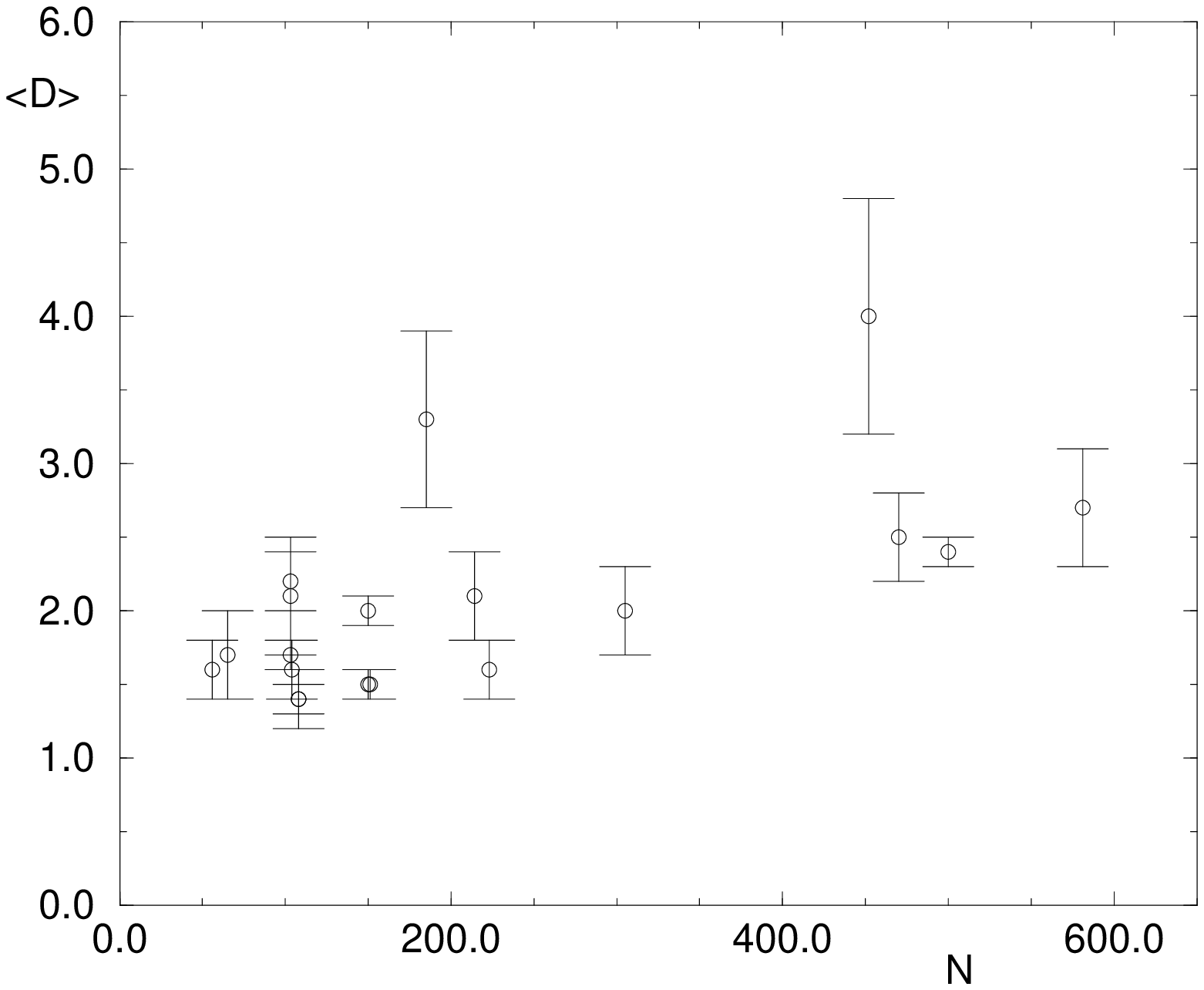,height=9.5cm,angle=0}}
\end{figure}
\newpage
\begin{figure}
\centerline{\psfig{figure=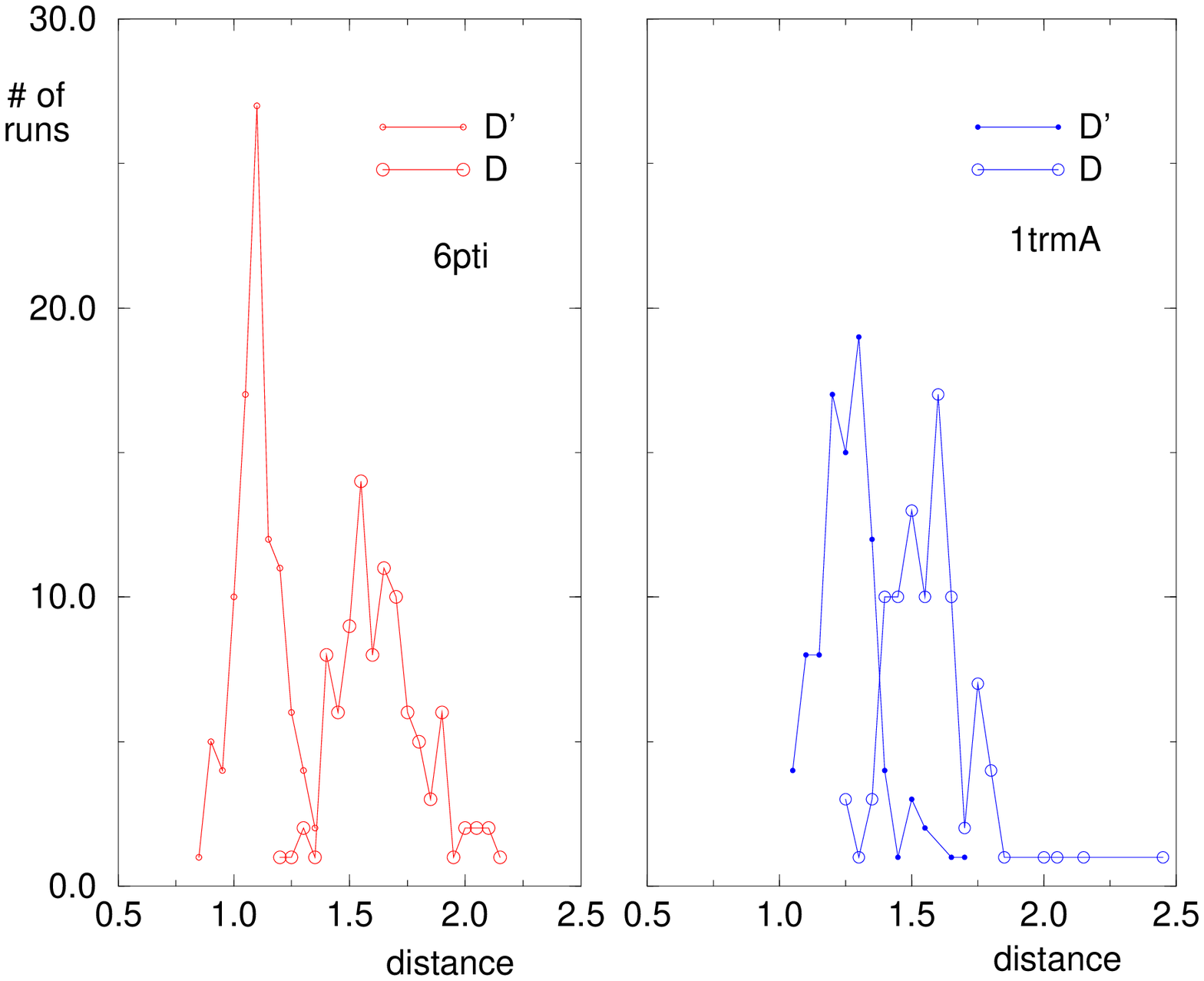,height=15cm,angle=0}}
\end{figure}
\newpage
\begin{figure}
\centerline{\psfig{figure=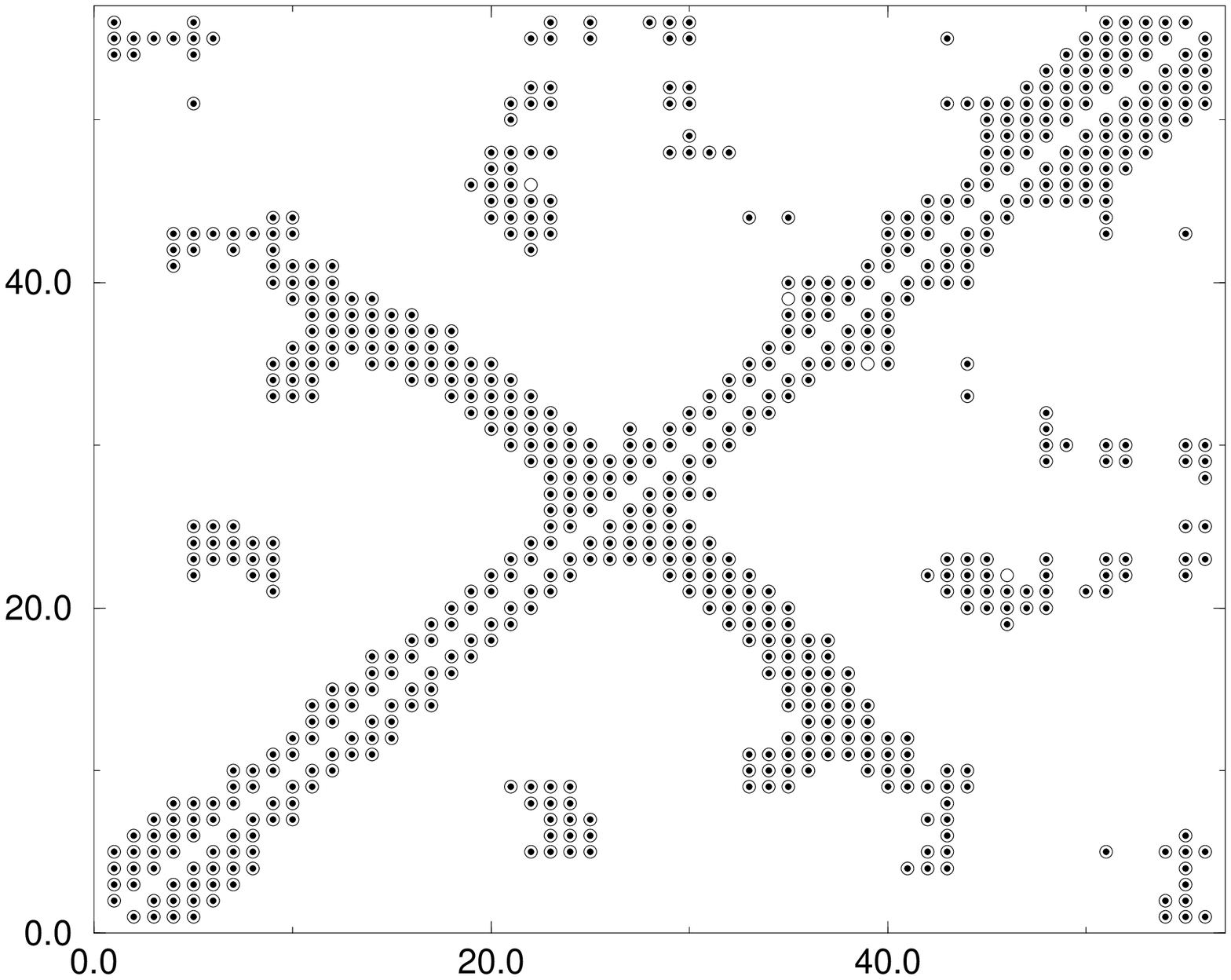,height=15.0cm,angle=270}}
\end{figure}
\newpage
\begin{figure}
\centerline{\psfig{figure=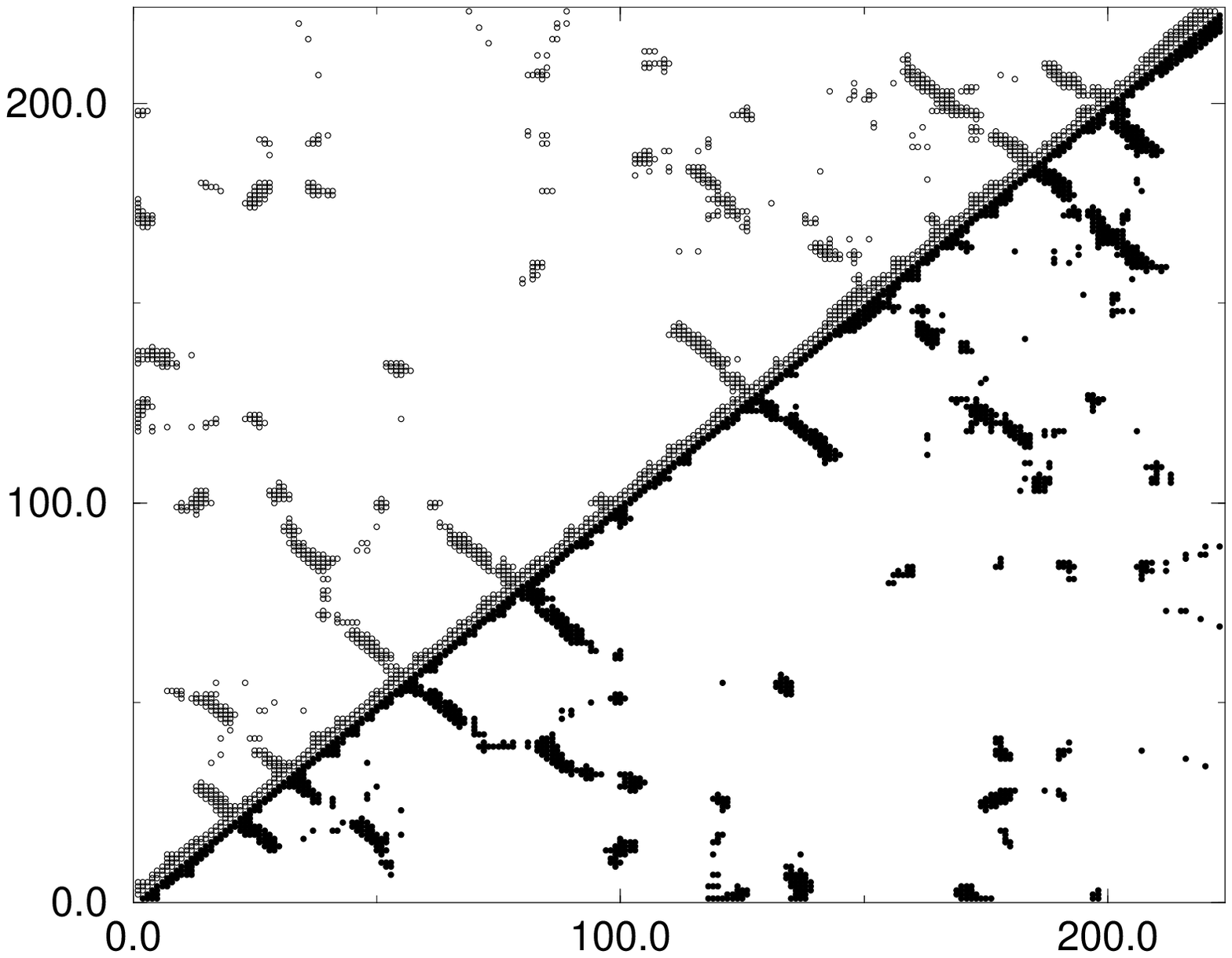,height=15.0cm,angle=0}}
\end{figure}
\newpage
\begin{figure}
\centerline{\psfig{figure=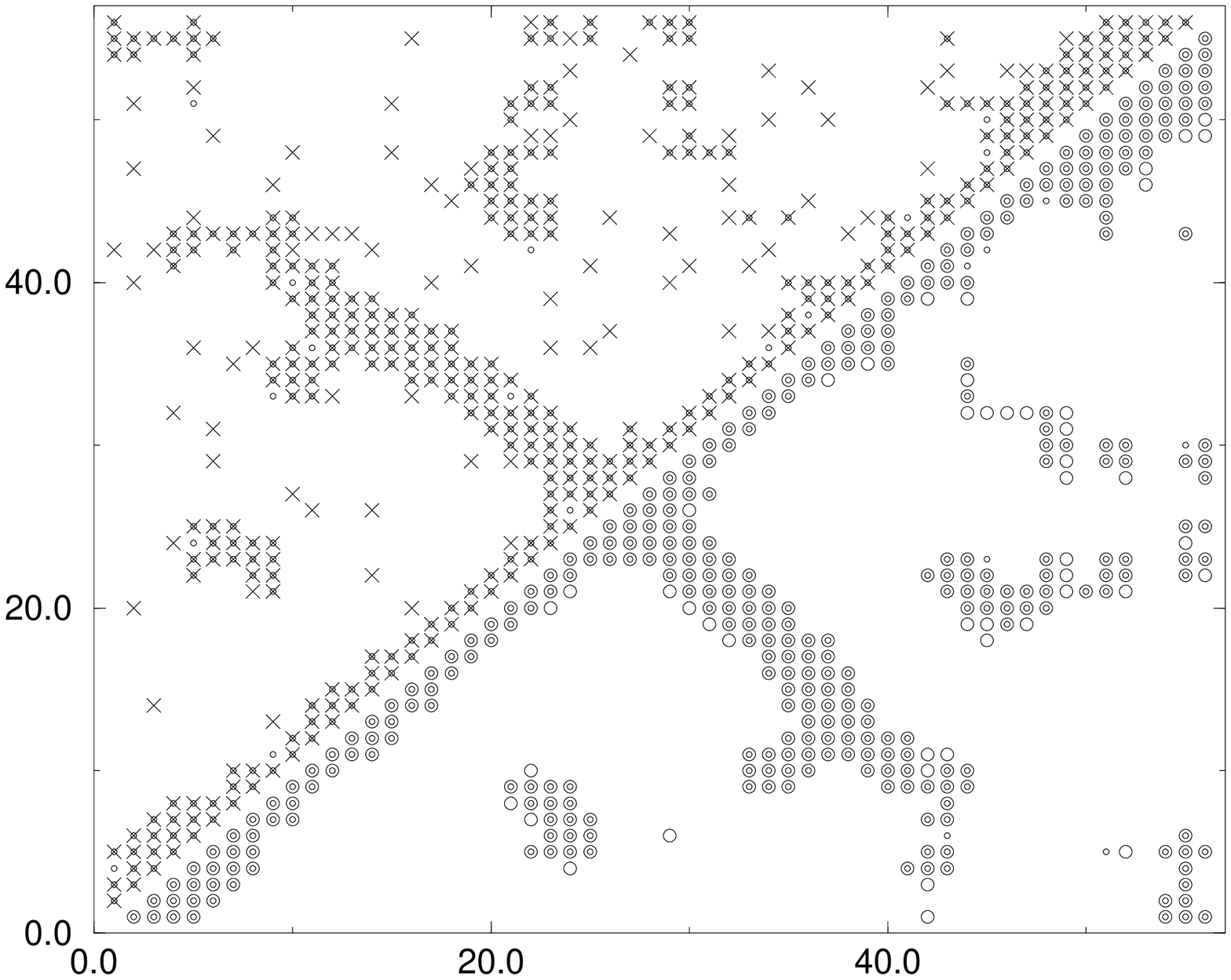,height=15.0cm,angle=270}}
\end{figure}
\newpage
\begin{figure}
\centerline{\psfig{figure=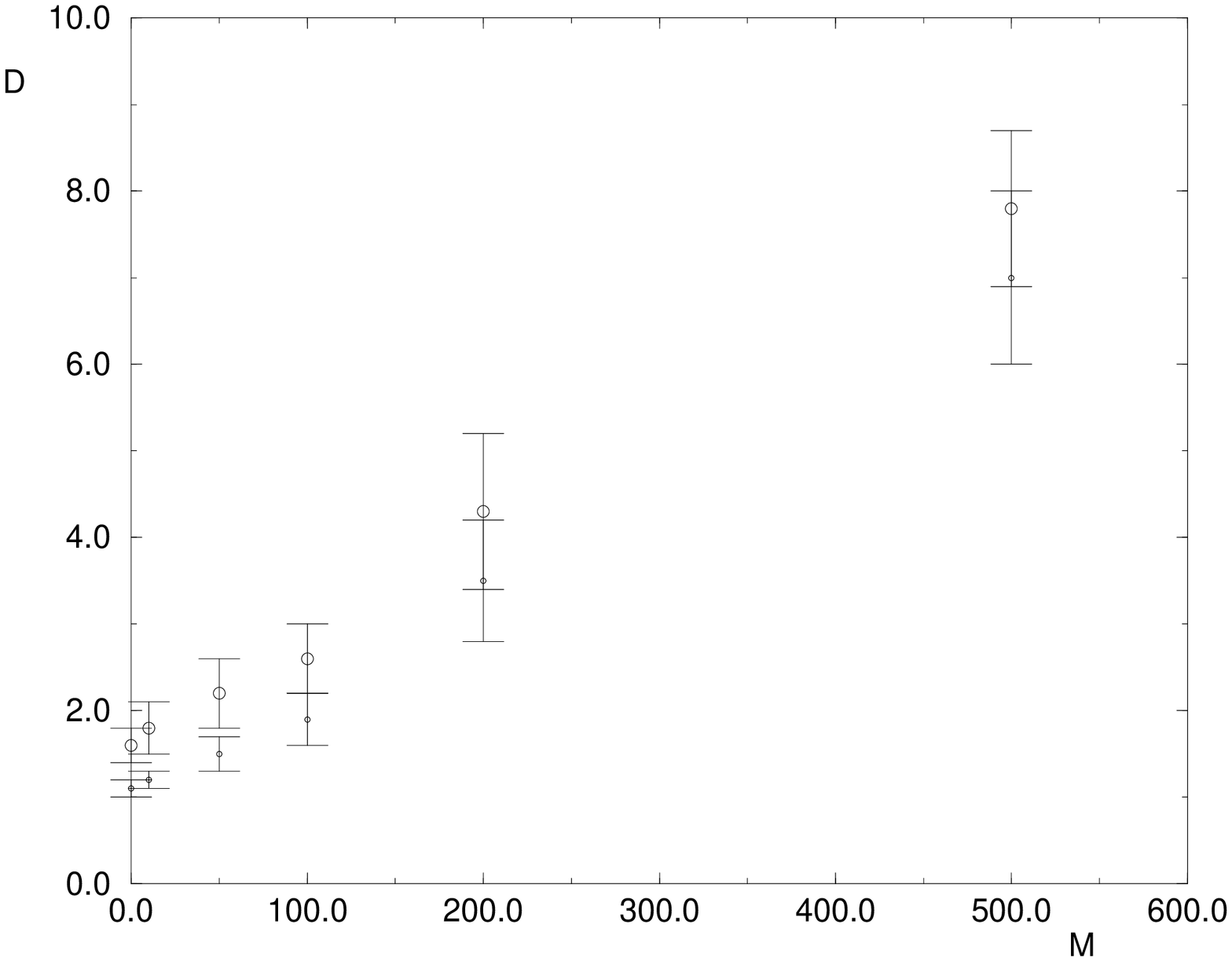,height=15.0cm,angle=270}}
\end{figure}
\newpage
\begin{figure}
\centerline{\psfig{figure=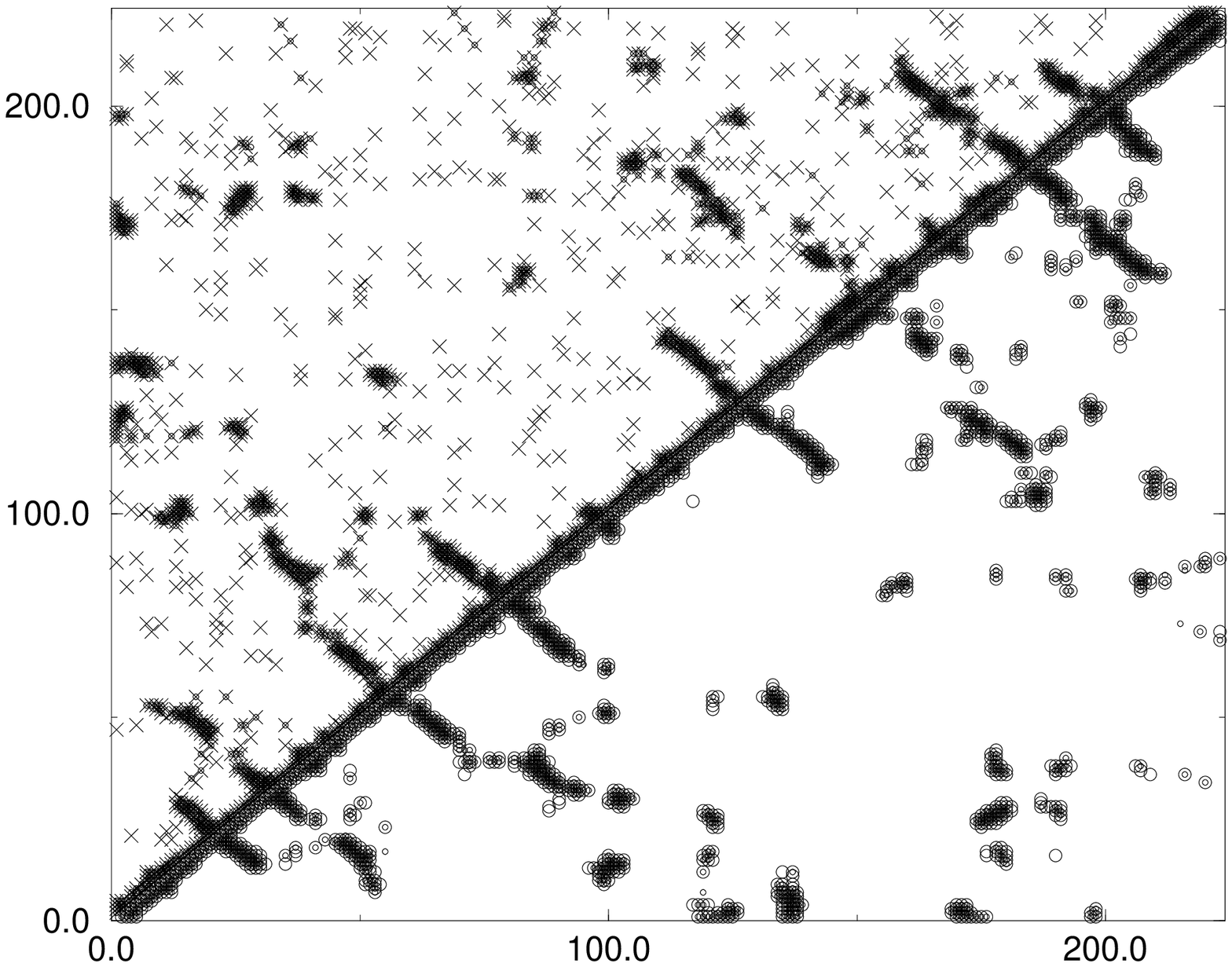,height=15.0cm,angle=270}}
\end{figure}
\newpage
\begin{figure}
\centerline{\psfig{figure=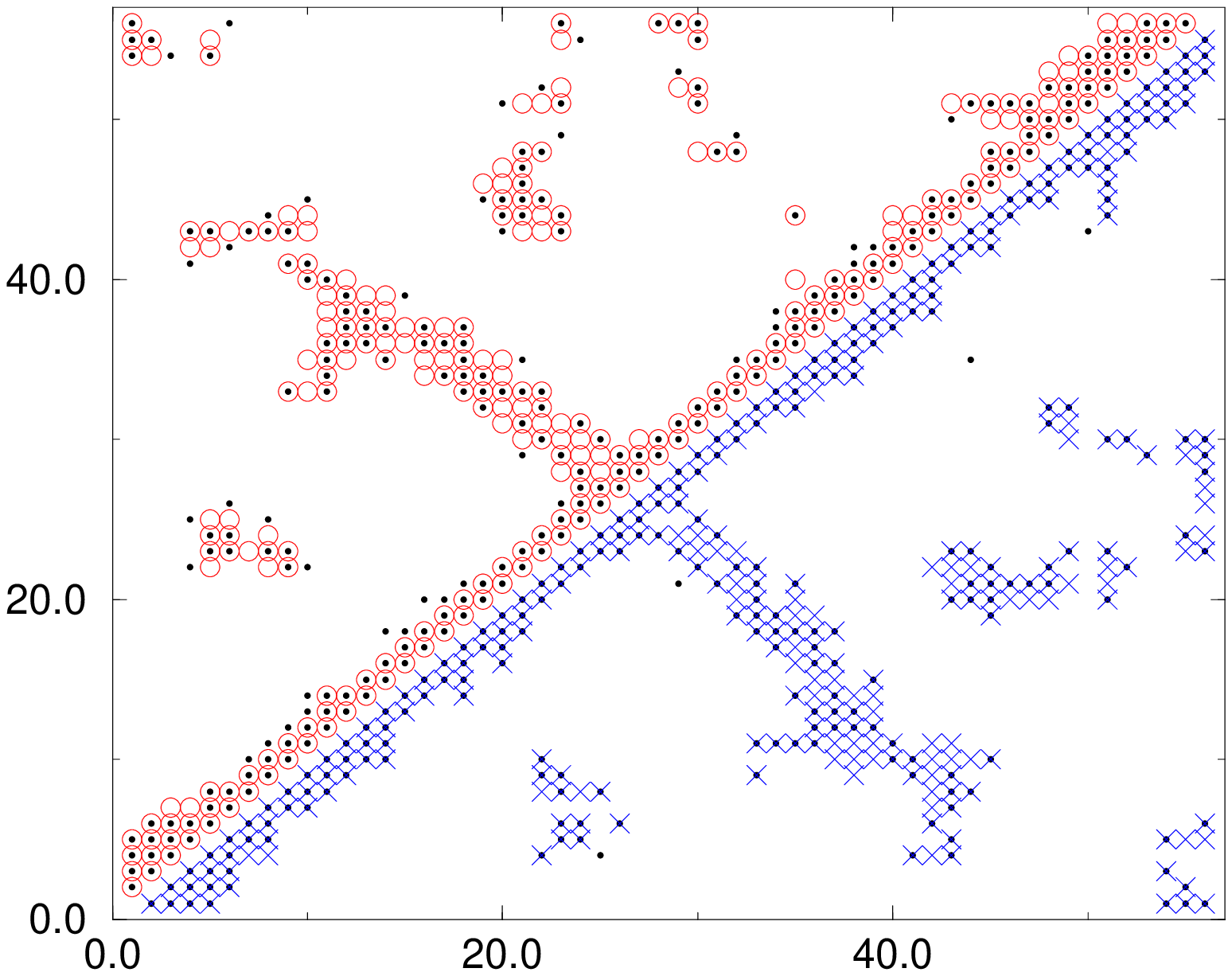,height=15.0cm,angle=0}}
\end{figure}
\newpage
\begin{figure}
\centerline{\psfig{figure=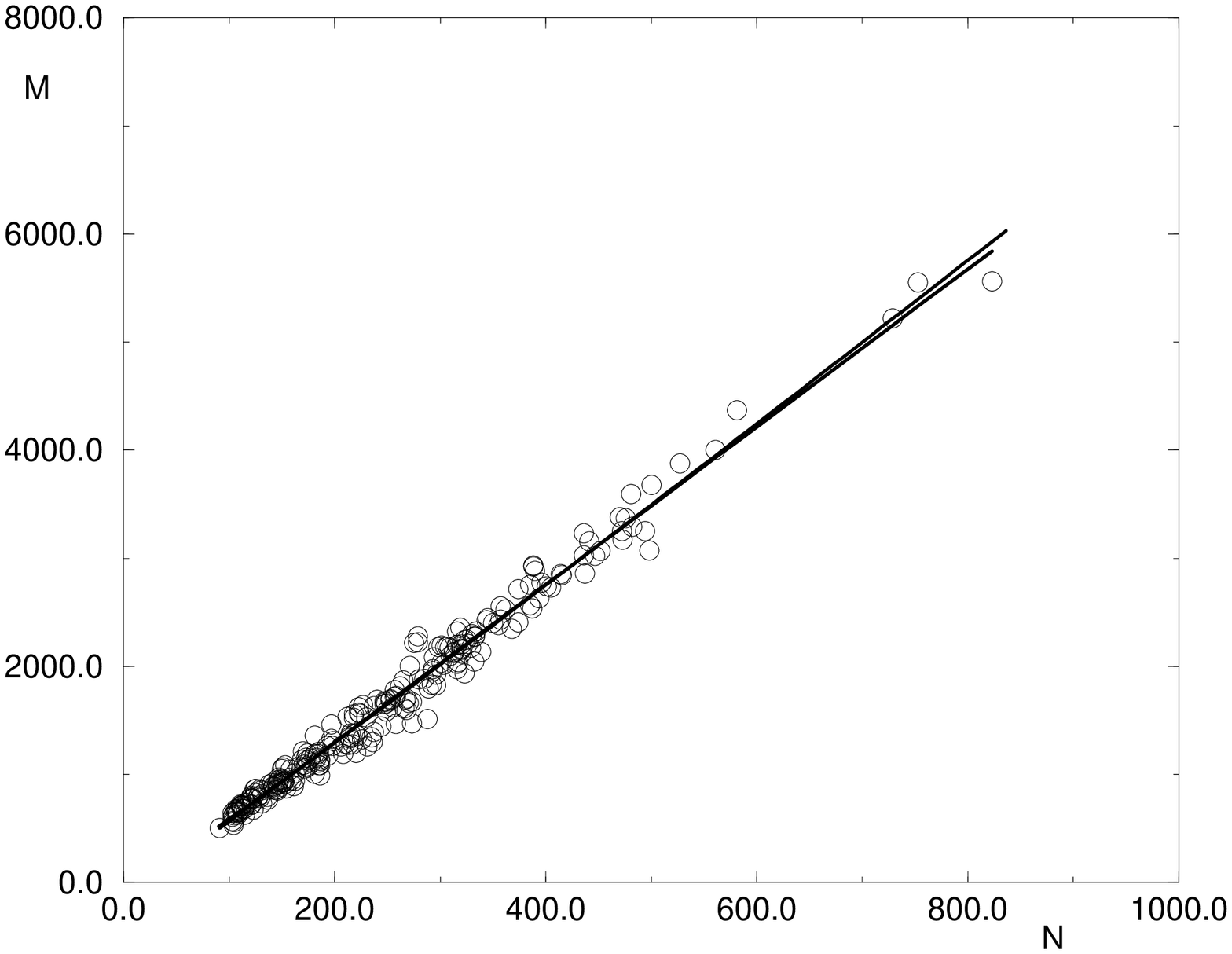,height=15.0cm,angle=270}}
\end{figure}
\newpage
\begin{figure}
\centerline{\psfig{figure=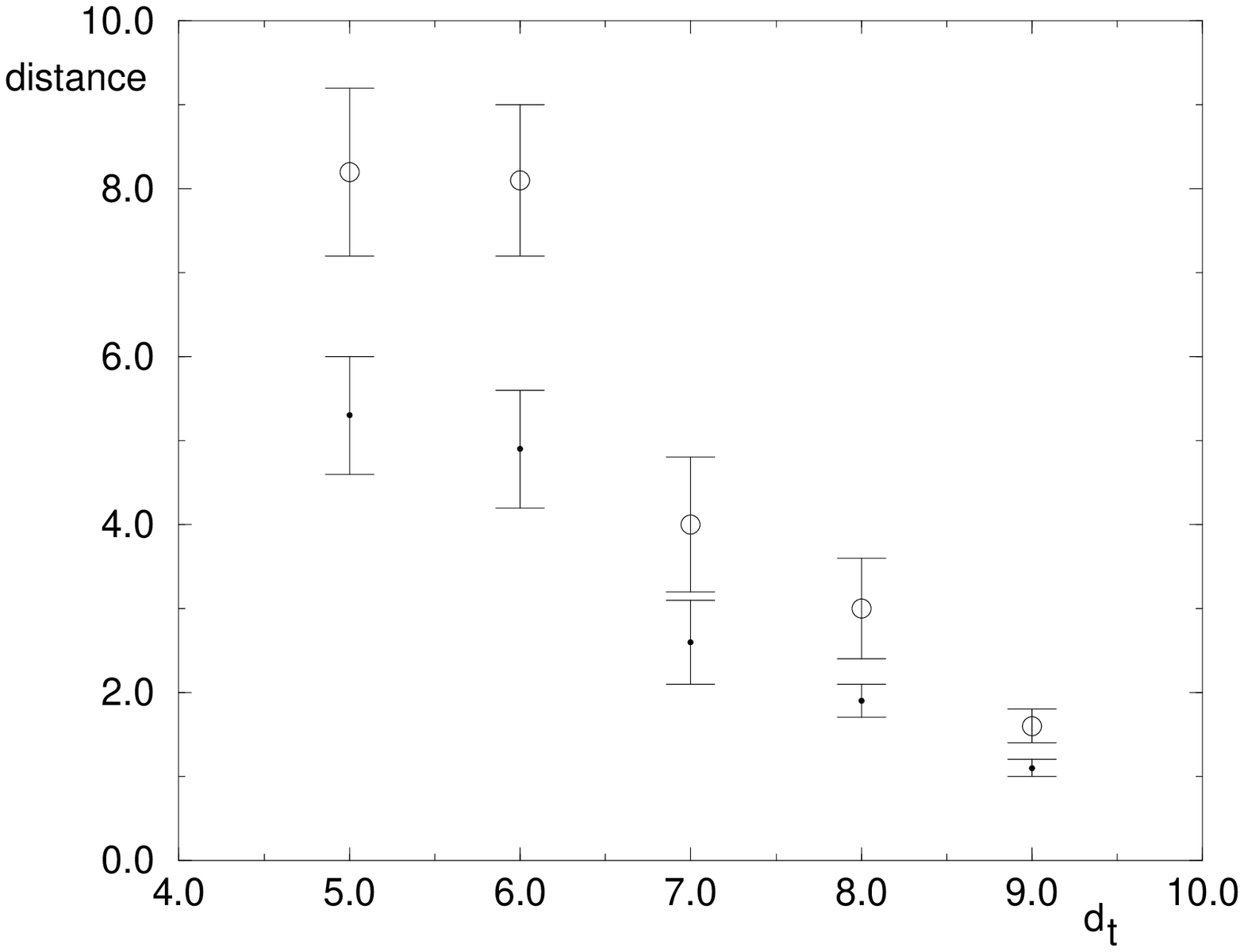,height=15.0cm,angle=0}}
\end{figure}


\newpage

\begin{tabular}{|ccccccc|}
\hline
\hline
protein & $N$ & $N_c$ & & protein & $N$ & $N_c$  \\
\hline
\hline
6pti    & 56  & 342  & & 2sodO   & 151 & 1066   \\
2ci2    & 65  & 350  & & 1hge1   & 185 & 1084   \\
1tlk    & 103 & 610  & & 1akeA   & 214 & 1348   \\
5cytR   & 103 & 644  & & 1trmA   & 223 & 1595   \\
1ltsD   & 103 & 571  & & 1abe    & 305 & 2179   \\
9rnt    & 104 & 623  & & 1pii    & 452 & 3070   \\
1acx    & 108 & 652  & & 3gly    & 470 & 3383   \\
2trxA   & 108 & 628  & & 3cox    & 500 & 3680   \\
1f3g    & 150 & 1049 & & 1gal    & 581 & 4369   \\
1aak    & 150 & 922  & &         &     &        \\
\hline
\hline
\end{tabular}

\newpage

\begin{tabular}{|c|c|c|c|c|c|}
\hline
\hline
A & B & C & AB & AC & BC  \\
\hline
 289 & 255 & 310  & 215 & 249 & 251  \\
\hline
\hline
\end{tabular}


\begin{references}
\bibitem{vd97}
Energy parameters are available upon request.
\end{references}
\end{document}